\documentclass[a4paper]{aa}
\usepackage{psfig,longtable,lscape}
\usepackage{epsfig}
\usepackage{graphicx}
\def \xmm {XMM--Newton}
\def \sax {BeppoSAX}
\def \src {SAX~J1748.2-2808}

\def \nh {N${\rm _H}$}

\def \hcm {\hbox {\ifmmode $ atom cm$^{-2}\else atom cm$^{-2}$\fi}}

\def \arcsec {\hbox{$^{\prime\prime}$}}

\def\approxgt{\mathrel{\hbox{\rlap{\lower.55ex \hbox {$\sim$}}
        \kern-.3em \raise.4ex \hbox{$>$}}}}
\def\approxlt{\mathrel{\hbox{\rlap{\lower.55ex \hbox {$\sim$}}
        \kern-.3em \raise.4ex \hbox{$<$}}}}

\begin{document}
\title{Unveiling the nature of the highly absorbed X--ray source SAX~J1748.2-2808 with XMM-Newton}


   \author{L. Sidoli\inst{1}
        \and S. Mereghetti\inst{1}
        \and F. Favata \inst{2}
    \and T. Oosterbroek \inst{3}
        \and A.N. Parmar \inst{2}}

\offprints{L.Sidoli (sidoli@iasf-milano.inaf.it)}

\institute{
        Istituto di Astrofisica Spaziale e Fisica Cosmica --
        Sezione di Milano -- IASF/INAF, I-20133 Milano, Italy 
\and
       Research and Scientific Support Department of ESA, ESTEC,
       Postbus 299, NL-2200 AG Noordwijk, The Netherlands
    \and
       Science Payload and Advanced Concepts Office, ESA, ESTEC,
       Postbus 299, NL-2200 AG, Noordwijk, The Netherlands
}

\date{Received 12 April 2006; Accepted: 6 June 2006}

\authorrunning{L. Sidoli et al.}

\titlerunning{{Unveiling the nature of SAX~J1748.2-2808 with XMM-Newton}}

\abstract
{We report on the results of an EPIC XMM-Newton observation of the faint
source SAX~J1748.2$-$2808 and the surrounding field. This 
source was discovered during the BeppoSAX Galactic center survey
performed in 1997-1998. 
A spatial analysis resulted in the detection of 31 sources within the EPIC field
of view. SAX J1748.2$-$2808 is clearly resolved into 2 sources in
EPIC images with the brighter contributing almost 80\% of the 2--10~keV flux. 
Spectral fits to this main source are
consistent with an absorbed power-law with a photon index of 
$1.4\pm 0.5$ and absorption equivalent 
to $14 ^{+6} _{-4}\times 10^{22}$~cm$^{-2}$ together with an iron line 
at $6.6 ^{+0.2} _{-0.2}$~keV with an equivalent 
width of $780 ^{+620}_{-380}$~eV. 
The significantly better statistics of the \xmm\ observation, compared with \sax,
allows to exclude a thermal nature for the X--ray emission. 
A comparison with other observations of SAXJ1748.2$-$2808 
does not reveal any evidence for spectral or
intensity long-term variability. 
Based on these properties we propose that
the source is a low-luminosity
high-mass X-ray binary located in the Galactic center region.
}
\maketitle
%

\section{Introduction}
\label{sect:intro}

\src\ is an X--ray source discovered with the Narrow Field
Instruments on-board \sax\ 
during a survey of the Galactic Center region
(hereafter GC) performed in September 1997 (Sidoli~\cite{s:00},
Sidoli et al.~\cite{s:01}).
The \sax\ spectrum was severely absorbed 
and displayed an intense Fe~K emission line. The spectrum
was poorly constrained, making both thermal and non-thermal nature
for the X--ray emission possible.
The source, unresolved at the angular resolution of the MECS
instrument  (FWHM~$\sim$1$'$), is located in the direction of the
giant molecular cloud Sgr~D. 

At the time of its discovery, the nature of {\mbox \src} was
uncertain and its intense Fe~K line emission, together with its
highly absorbed spectrum, made it a unique object in the GC region
which could well represent the bright tail of a distribution of
similar unresolved objects significantly contributing to the diffuse Fe line
emission (at 6.7~keV) from the galactic ridge (Koyama et
al.~\cite{k:89}; Ebisawa et al.~\cite{e:01}).

Interestingly, \src\ displays properties very similar to a class
of sources subsequently discovered with the INTEGRAL satellite
(see e.g., Walter et al.~\cite{w:04}, and references therein):
these objects show strong photoelectric absorption, hard 2--10 keV
spectra,  and often display intense Fe line emission. Most of them
also show X--ray pulsations, thus indicating that they are likely
high mass X--ray binaries embedded in a local absorbing gas.

Here we report the results of an \xmm\ observation the region of
sky around \src, performed with the main goal of unveiling the
nature of this intriguing source.

 \begin{figure*}[ht!]
  \centering
   \includegraphics[angle=0,height=8cm]{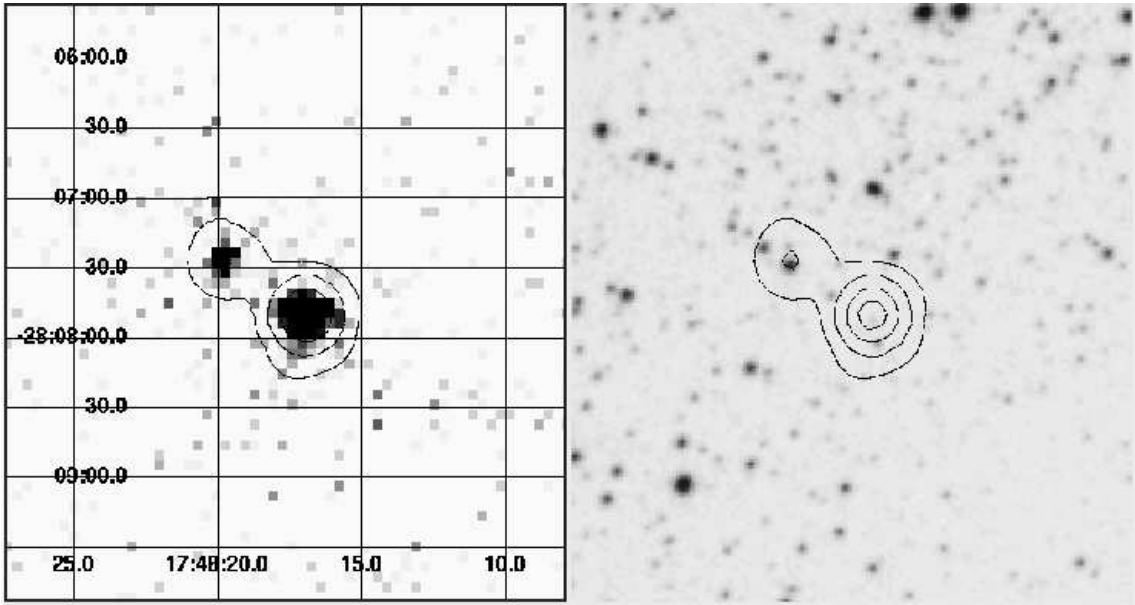}
      \caption{{\em Left panel:} Combined (pn+MOS1+MOS2) EPIC 
image (2--10 keV) 
centered on \src. Contours (at 5, 10, 20 and 30 counts/pixel) 
mark the
two sources (the ``main'' and the ``faint'') resolved with \xmm. 
{\em Right panel:} Optical image 
of the source field, from the ``Second Epoch Survey" of the southern 
sky made by the Anglo-Australian Observatory (AAO) with the UK Schmidt Telescope
(digitized plates available from  STScI at {\em http://archive.stsci.edu/}).
The star positionally coincident with the faint source is 0600-28834001 in the USNO-A2.0 catalog.}
\label{fig:resolved}
   \end{figure*}

\section{Observations}
\label{sect:obs}

The XMM-Newton Observatory (Jansen et al. \cite{ja:01}) includes
three 1500~cm$^2$ X--ray telescopes each with an European Photon
Imaging Camera (EPIC) at the focus, composed of and one pn (Str\"uder et al.
\cite{st:01}) two MOS CCD  detectors (Turner et al.~\cite{t:01}). 
The \src\ field was observed with \xmm\ on 2005 February
26-27 for about 50~ks.

Data were processed using version 6.1 of the \xmm\ Science
Analysis Software (SAS). Known hot, or flickering, pixels and
electronic noise were rejected using the SAS. A further severe
cleaning was necessary because of the presence of several soft
proton flares. After rejecting the time intervals where the flares
were present, the net good exposure times reduced to about 32.3~ks
for the MOS1 and the MOS2, and to 13.3 for the pn. 
Cleaned MOS1
(with pattern selection from 0 to 12), MOS2 (pattern selection
from 0 to 12) and pn (patterns from 0 to 4) events files were
extracted, and used for the subsequent analysis: a source
detection analysis and a spectral analysis of the brightest
sources.

Spectra were rebinned such that at least
20 counts per bin were present and such that the energy resolution
was not over-sampled by more than a factor 3. Free relative
normalizations between the MOS1, MOS2 and pn instruments were
included. The background spectra were extracted from source free
regions of the same observation. All spectral uncertainties and
upper-limits are given at 90\% confidence for one interesting
parameter.

\section{\src: analysis and results}

A close-up view of the combined EPIC 2--10 keV
image is shown in Fig.~\ref{fig:resolved}, together with an optical
image of the same field. 
\xmm\ resolved \src\ into
two sources, a brighter (here called ``main'' source)
and a fainter one.

In order to minimize contamination from the nearby fainter source,
we extracted source counts centered on the ``main'' source, from a
circular region of 20\arcsec\ radius, for the MOS1, MOS2 and pn
detectors separately.

\begin{figure}[ht!]
  \centering
   \includegraphics[angle=-90,width=8.0cm]{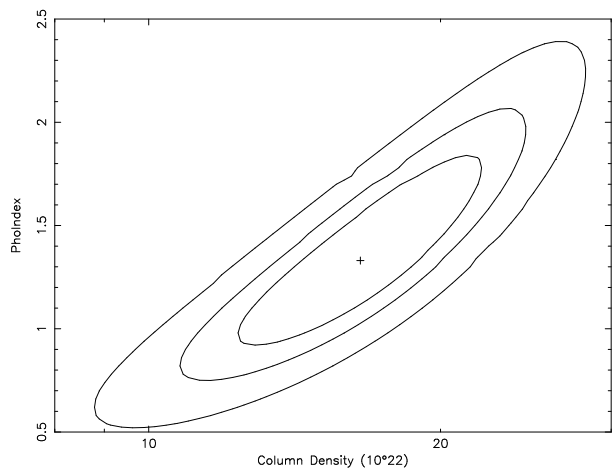}
      \caption{68\%, 90\% and 99\% confidence level contours for the
two quantities derived with the absorbed power-law fit
to the ``main'' source spectrum,
absorbing column density and power-law photon index.}
         \label{fig:cont_pow}
  \end{figure}

\begin{figure*}
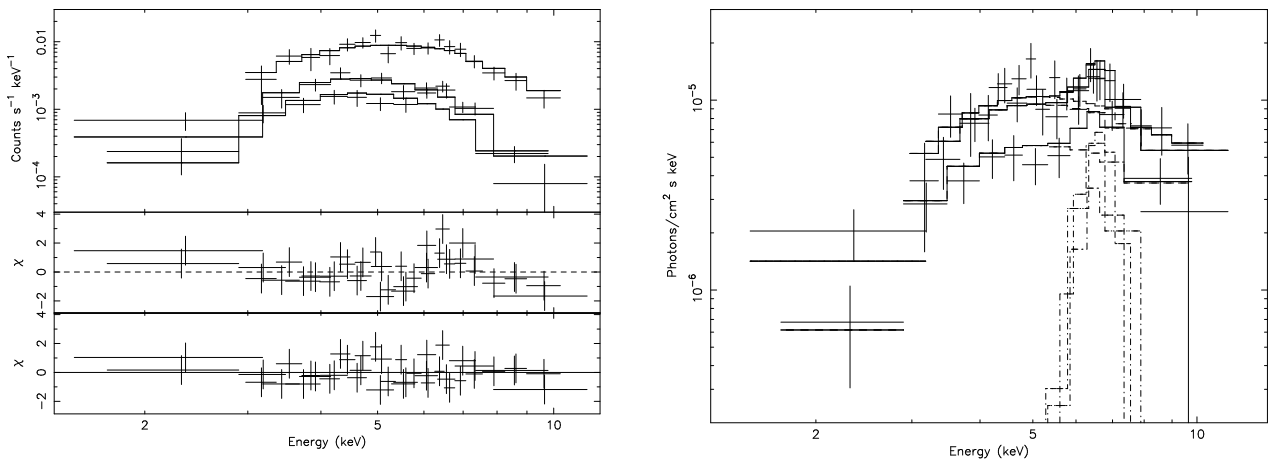

\hbox{\hspace{0.5cm}
\includegraphics[height=7.9cm,angle=-90]{5418fig3.ps}
\vspace{1.0cm}
\hspace{.6cm}
\includegraphics[width=6.0cm,angle=-90]{5418fig4.ps}}
\vbox{\vspace{-3.4cm}}
\hbox{\hspace{.6cm}
\includegraphics[height=7.8cm,angle=-90]{5418fig5.ps}}

\caption[]{{\em Left panel:} Fit to the ``main'' source spectrum (MOS1, MOS2 and pn data)
with a single absorbed power-law ({\em top panel}) together with its residuals
in units of standard deviations ({\em middle panel}). There is evidence for 
an excess around 6--7~keV, indicative of the presence of an emission line. 
The residuals 
after adding a Gaussian line to the power-law model are shown in the {\em bottom panel}, in units
of standard deviations. 
{\em Right panel}: Photon spectrum resulting from the best-fit, 
composed by a power-law together with
an emission line at 6.6~keV (see Table~\ref{tab:mainspec} for the parameters).
}
\label{fig:mainspec}
\end{figure*}

The fit of the spectrum with an absorbed power-law 
($\chi^2$=41.5 with 35 degrees of freedom, d.o.f.) resulted in 
structured residuals around 6--7~keV, confirming the presence of the Fe-K line
already observed with \sax.
Adding a Gaussian line to the power-law model resulted in a 
significantly better fit ($\chi$/d.o.f.=24.9/32), with a broad
line ($\sigma$=0.43$^{+0.33}_{-0.20}$~keV) centered in the
range 6.4--6.8 keV (see Figs.~\ref{fig:cont_pow} and ~\ref{fig:mainspec}, and
Table~\ref{tab:mainspec} for the resulting parameters).
We next tried with other simple models, such as a
hot plasma model ({\sc mekal} in {\sc xspec}), 
a thermal bremsstrahlung spectrum, and a blackbody model (see
Table~\ref{tab:mainspec} for the spectral analysis results).

\begin{table*}[!ht]
\vspace{0.5cm}
\begin{center}
\caption[]{Results obtained fitting the ``main'' source spectrum.
The 2--10 keV flux has not been corrected for the interstellar
absorption. The luminosity, $L$, has been obtained from the
unabsorbed flux, assuming the GC distance (8.5~kpc). E$_{line}$ is
the Gaussian line centroid in keV, and EW is the associated
equivalent width (keV),  $\sigma$ is the line width in keV, while I$_{line}$ is the total
10$^{-6}$~photons~cm$^{-2}$~s$^{-1}$ in the line. }
\label{tab:mainspec}
\begin{tabular}{llllll}
\hline
Model          & Column density     & Parameter                          & $\chi^2$/dof  &Flux  (2--10 keV)                  & $L$ (2--10 keV)   \\
               &($10^{22}$ cm$^{-2}$)&                                    &           &($10^{-13}$ erg  cm$^{-2}$ s$^{-1}$ ) & ($10^{34}$ erg s$^{-1}$ )\\
\hline
Power law      & $16.5^{+6.0}_{-4.5} $       & $\Gamma=1.3^{+0.6}_{-0.3}$               & $41.5/35$ &     $6.8$     &  1.1    \\
Power law + line & $14.3^{+6.0}_{-4.0} $       & $\Gamma=1.4^{+0.4}_{-0.5}$             & $24.9/32$ &     $6.6$     &  1.0    \\
                 &                             & E$_{line}$=$6.6^{+0.2}_{-0.2}$         &           &                &         \\
                 &                             &  $\sigma$=0.43$^{+0.33}_{-0.20}$  &           &                &          \\
                 &                             & EW=$0.78^{+0.62}_{-0.38}$                 &           &                &          \\
                 &                             & I$_{line}$=$9.0^{+7.0}_{-4.1}$         &           &                &          \\
{\sc mekal}      &   $16.6^{+2.7}_{-2.7}$                & T$_{\rm M}=30^{+50}_{-15}$ keV   & $38.1/35$ &     $6.7$     &  1.1   \\
Bremsstrahlung   &   $16.8^{+4.8}_{-3.2}$      &T$_{\rm br}> 12$ keV                     & $41.3/35$ &     $6.8$     &  1.1   \\

Black body       &   $10.4^{+3.9}_{-3.2}$      &T$_{\rm bb}=2.2^{+0.4}_{-0.3}$ keV   & $37.5/35$      &    $6.7$ &  0.8 \\
                 &                             & R$_{\rm bb}$=$0.07^{+0.02}_{-0.02}$ km  &           &           &          \\

\noalign {\smallskip}
\hline
\end{tabular}
\end{center}
\end{table*}


The second fainter source resolved with \xmm\ is too weak to allow
a meaningful spectral analysis.

\begin{figure}[!ht]
  \centering
\hspace{-1.1cm}
   \includegraphics[angle=-90,width=7.5cm]{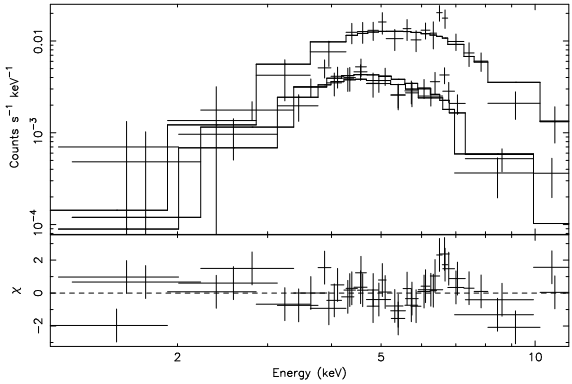}
\caption{\xmm\ counts spectrum, of the
``main'' plus ``faint'' sources, fitted using
an absorbed power-law (see text for the  parameters), together with the residuals
in units of standard deviations}
\label{fig:unres_res}
\end{figure}

\begin{figure}[!ht]
  \centering
\hspace{-1.1cm}
   \includegraphics[angle=-90,width=7.5cm]{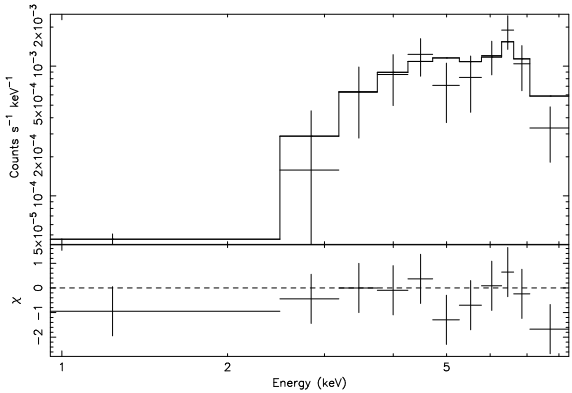}
\caption{\sax\ MECS2+MECS3 spectrum (data points) comprising 
the emission of both the ``main" and ``faint" sources, unresolved by this 
instrument. The line is the XMM-Newton best fit model to the sum of the 
two sources (Fig.~\ref{fig:unres_res}) folded through the BeppoSAX response.
}
\label{fig:sax}
\end{figure}

It is reasonable to assume that the \src\ emission was mostly
contributed by the brighter, ``main'', source (and in the
following we will call it \src). The fainter source probably
contributed a fraction of the measured flux from the iron line in
the \sax\ spectrum.
In order to allow a proper comparison with the \sax\ observation,
we extracted a combined \xmm\ spectrum from both the main and the
secondary fainter source. The residuals to the fit with an
absorbed power-law again clearly show an excess around
6.5--6.7~keV (see Fig.~\ref{fig:unres_res}), requiring the addition of a Gaussian emission line.
The line centroid is 6.6$^{+0.2}_{-0.1}$~keV, the normalization
is (8$^{+9}_{-3}$)~$\times$10$^{-6}$~photons~cm$^{-2}$~s$^{-1}$ and the
equivalent width is 400$^{+500} _{-150}$~eV. The absorbing column
density is (13$^{+7} _{-3}$)~$\times$10$^{22}$~cm$^{-2}$, while
the photon index, $\Gamma$, is 1.2~$^{+0.9}_{-0.3}$ (which is harder than the
power-law fit to the ``main'' source alone, likely due to the hard
emission contributed by the fainter source). The
observed flux is 9$\times$10$^{-13}$~erg~cm$^{-2}$~s$^{-1}$
(2--10~keV). We then re-extracted the \sax\ MECS spectrum, and
deconvolved it with this best fit model for the \xmm\ main plus
faint sources emission. The result is shown in Fig.~\ref{fig:sax}
and demonstrates that there is no evidence for dramatic changes in
the flux level and spectral shape.

\section{A source catalog of the \src\ field}
\label{sect:det}

Examination of the EPIC images of the \src\ field in different
energy ranges shows a region rich in faint X--ray sources.
We  performed a detection analysis in
order to obtain a source catalog of this region, using 
the source detection procedure described in Baldi
et al.~(\cite{b:02}). All the source detection chain uses SAS
version 6.1 tasks.

The cleaned events files were used to produce MOS1, MOS2 and pn
images and exposure maps (which also include the vignetting
effect) in 4 energy ranges: 0.5--2~keV (soft band, hereafter
``S''), 2--5~keV (medium band, ``M''), 5--10~keV (hard band,
``H''), and 0.5--10~keV (total band). For each energy
band independently, the MOS1, MOS2 and pn images (and the
corresponding instrumental exposure maps) were then merged in  
order to get a higher signal-to-noise ratio in the source
detection. After the production of a detector mask (with the task
{\em emask}), a source detection in local mode was performed
in each energy band separately with the task {\em eboxdetect} to
produce a preliminary list of sources using a sliding box
technique. Then, with task {\em esplinemap}, all the sources in
the list were removed from the image and the resulting
source--free image was fitted with a cubic spline function in
order to create a background map for each energy band. Then, a
second run with {\em eboxdetect} in map mode was made, this
time using the background maps produced before. Lastly, the final
source positions, together with the EPIC combined (MOS1+MOS2+pn)
count rates for each source in each energy range, were calculated
using the task {\em emldetect}, which performs maximum likelihood
fits to the source spatial count distribution in all energy bands.

This detection procedure resulted in 31
sources, reported in Table~\ref{tab:cat}. Most of them
have been detected in the medium and hard energy ranges, while few
sources were detected only in the soft band (0.5-2~keV), very
likely foreground stars.

The observed fluxes have been calculated from the combined EPIC count rate, $cr$,
and a total conversion factor $f$ (flux=$f$ $\cdot$~$cr$)
derived as in Baldi et al.~(\cite{b:02}):

$$
\frac{T_{tot}}{f}=\frac{T_{MOS1}}{f_{MOS1}}+\frac{T_{MOS2}}{f_{MOS2}}+\frac{T_{pn}}{f_{pn}}\:,
$$
where $T_{tot}$ is the sum of the exposure times from the three
instruments, $T_{tot}$=$(T_{MOS1}+T_{MOS2}+T_{pn}$), and
$f_{MOS1}$, $f_{MOS2}$ and $f_{pn}$ are the single
count-rate-to-flux conversion factors.

\begin{figure*}[!ht]
\hbox{\hspace{.0cm}
\includegraphics[height=6.0cm,angle=0]{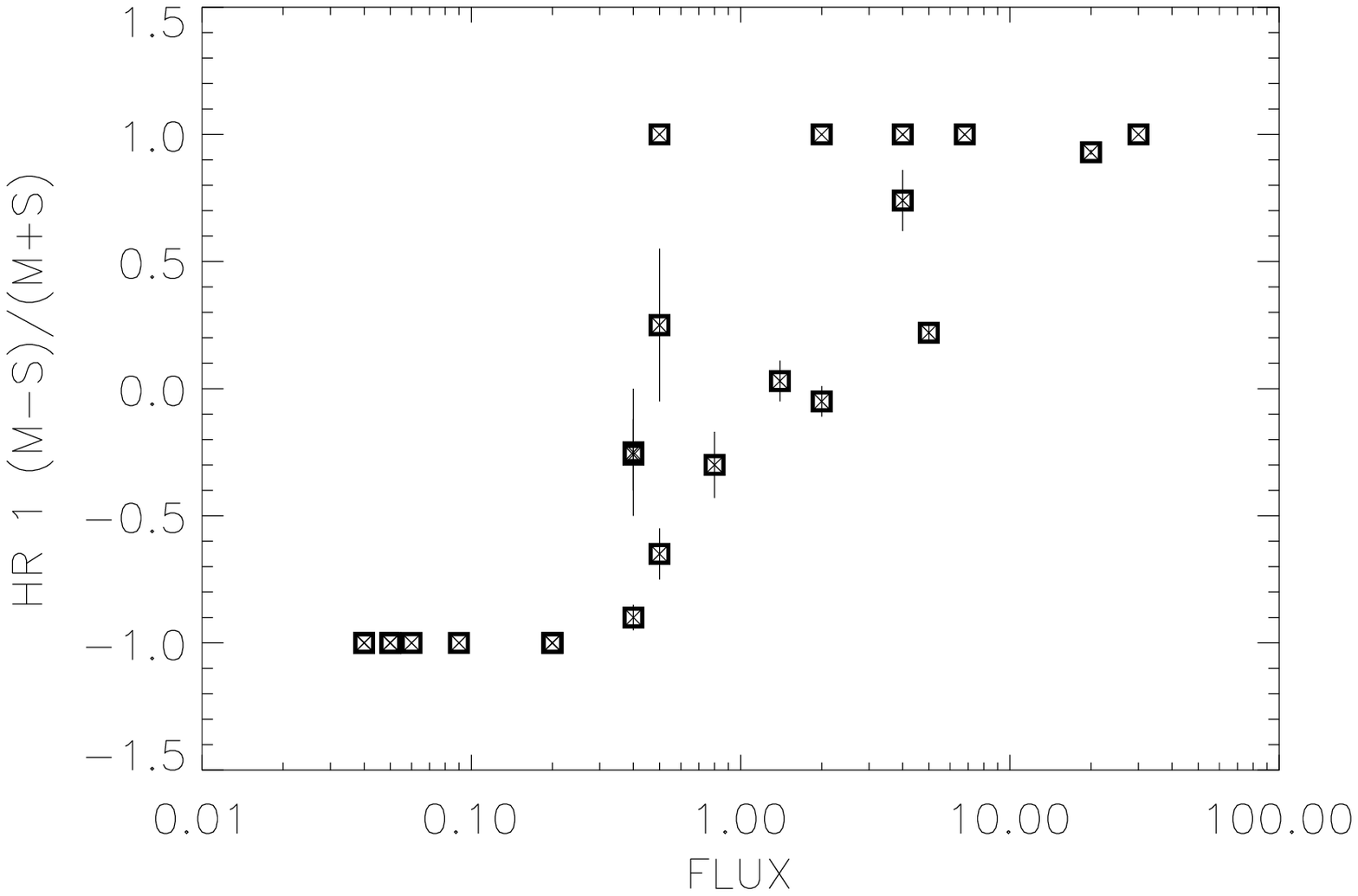}
\hspace{0.9cm}
\includegraphics[height=6.cm,angle=0]{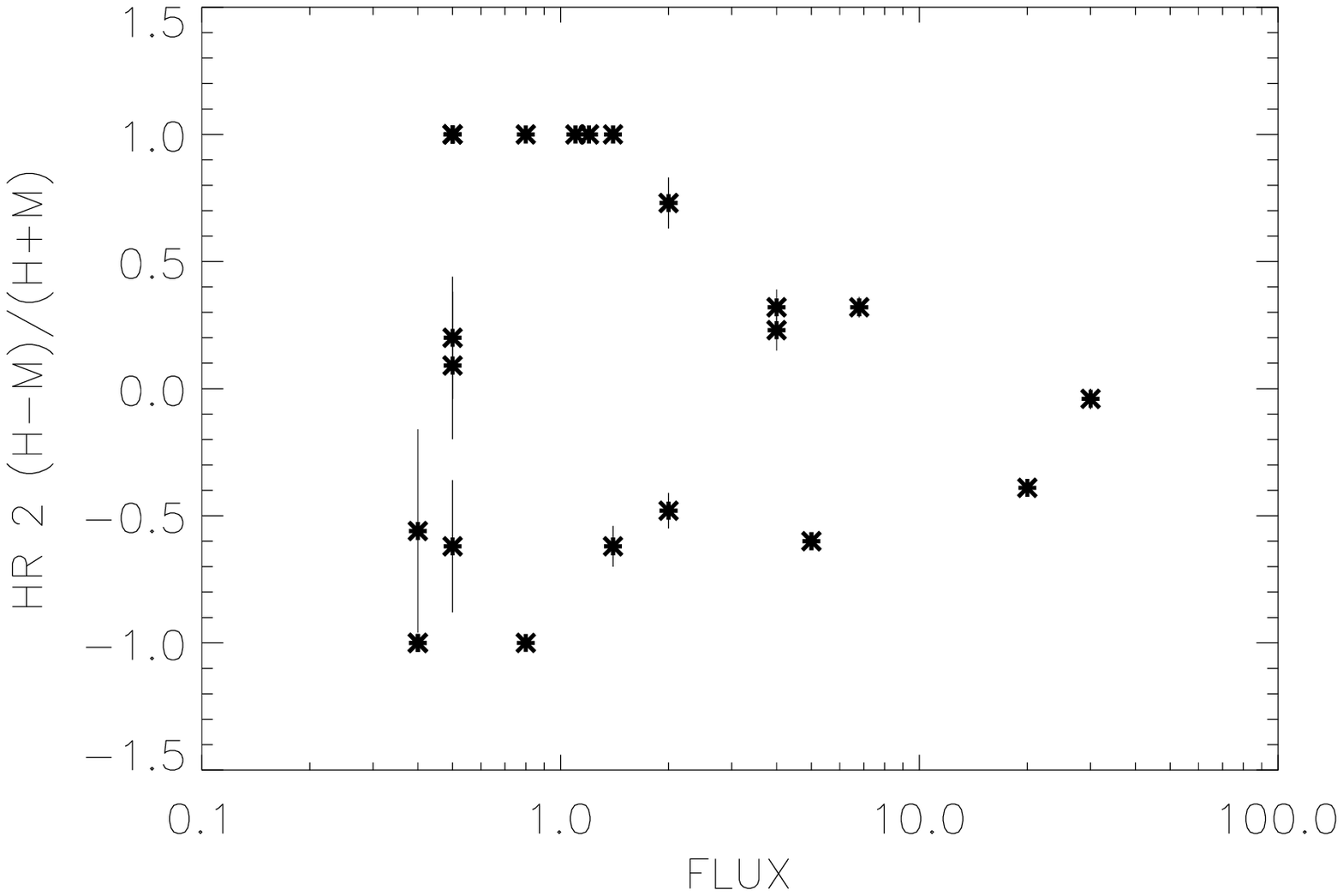}}
\caption[]{Hardness ratio versus estimated flux in the energy
range 2--10~keV for the X--ray sources in the \src\ field. On the
left the hardness ratio HR1 ((M--S)/(M+S)) versus flux is shown,
while in the right panel the distribution of the hardness ratio
HR2 ((H--M)/(H+M)) versus observed flux is reported. Fluxes are in
units of 10$^{-13}$~erg~cm$^{-2}$~s$^{-1}$ (see
Table~\ref{tab:cat}). } \label{fig:hr}
\end{figure*}

The sources listed in our catalog show a distribution of hardness
ratios, ranging from sources detected only in the soft band
(likely foreground stars), up to extremely hard sources, detected
only above 5~keV (see Fig.~\ref{fig:hr}). 
In order to account for the different source
hardness, we derived three different total conversion factors
$f$, assuming three kinds of ``typical'' spectra: 
for sources detected only in the ``S'' band, we considered a
blackbody spectrum with temperature 200~eV and a column
density of $10^{21}$~cm$^{-2}$ (and the derived flux
is limited to the 0.5--2~keV energy range); for sources detected only in the ``H'' band,
we have chosen a power-law spectrum with a photon index $\Gamma$=2  and an absorbing
column density of $10^{24}$ cm$^{-2}$
(flux derived in the 5--10 keV energy range); 
a power-law spectrum with a photon index $\Gamma$=2
absorbed with \nh=2$\times$$10^{23}$ cm$^{-2}$
has been considered for all other sources (flux derived in the 2--10 keV range).

The absorbed fluxes calculated in this way are reported in
Table~\ref{tab:cat} for the faint sources, while for the brightest
ones, for which a reliable spectrum can be obtained, the flux
measured from the spectral analysis is reported. 

\begin{table}[!ht]
\vspace{0.5cm}
\begin{center}
\caption[]{X--ray spectral results for the brightest sources
detected in the \src\ field (the source ID is the same as in
Table~\ref{tab:cat}). The flux has not been corrected for
interstellar absorption and is in units of $10^{-13}$ erg~cm$^{-2}$~s$^{-1}$ 
in the energy range 2--10~keV, except for
source~6, which is in the 0.5--2~keV range. Al the source
spectra have been fit with a power-law, except for source~6,
where a blackbody has been used (the blackbody temperature,
T$_{\rm bb}$, is in keV). } \label{tab:other}
\begin{tabular}{llll}
\hline
Src ID          & \nh     & Parameter                          & Flux       \\
               &($10^{22}$ cm$^{-2}$)&                                    &  (2--10~keV)         \\
\hline

1               & $1.3^{+0.3}_{-0.3} $  & $\Gamma = 1.7^{+0.2}_{-0.1}$   &    5        \\
2               & $8^{+2}_{-2} $  & $\Gamma = 2.0^{+0.4}_{-0.6}$   &      20       \\
4               & $15^{+5}_{-6} $  & $\Gamma = 1.6^{+0.9}_{-0.9}$   &   30        \\
5               & $0.7^{+0.8}_{-0.4} $  & $\Gamma = 1.8^{+0.8}_{-0.6}$   &  2          \\
6               & $<0.2$              &  T$_{\rm bb}$=$0.2\pm{0.1}$ &   0.5   (0.5--2~keV)        \\
7               & $1.6^{+1.3}_{-0.7} $  & $\Gamma=2.1^{+1.2}_{-0.6}$   &    1.4        \\
\noalign {\smallskip}
\hline
\end{tabular}
\end{center}
\end{table}


A search in the SIMBAD database resulted in two HD stars
positionally coincident with two soft X-ray sources (see
Table~\ref{tab:cat}): HD~316290, located  3\farcs1 from source~11, 
and HD~161824, at 1$''$ from the soft X--ray source~15. A
third star, Tyc2~929, is located within 1\farcs2 from the soft
X--ray source~28. In Table~\ref{tab:stars} we report the sky
positions and B, V magnitudes of these 3 optical counterparts,
together with the associated ratios between X--ray and optical
flux (Maccacaro et al. 1988), which strengthen the physical association 
with the soft X--ray sources.

\begin{table*}[ht!]
\caption{Stars positionally coincident with 3 soft X--ray sources
in our catalogue. } \label{tab:stars}

\begin{tabular}[c]{llllllll}
\hline\noalign{\smallskip}
Star  & R.A. (J2000)     & Dec. (J2000)         & B      &  V     &  Sp. type & Associated    & log(f$_{X}$/f$_{V}$)    \\
      &                  &                      & (mag)  &  (mag) &       & X--ray source ID  &     \\
\noalign{\smallskip\hrule\smallskip}
HD~316290     &  267.203000 & $-$28.016972      & 10.25   & 9.76    & F8          & 11  &  $-4.3$ \\
HD~161824     &  267.212250 & $-$28.246139      &  9.63   & 8.33    & K1/K2III    & 15  &  $-4.5$ \\
Tyc2 929      &  267.045025 & $-$28.308665      & 12.27   & 10.55   &             & 28  &  $-4.8$ \\
\noalign{\smallskip\hrule\smallskip}
\end{tabular}
\end{table*}

In the search for counterparts of the X--ray sources at other
wavelengths, we have conservatively assumed a circular uncertainty
region with a radius of 4$''$. The results from a search in the
SIMBAD database are reported in Table~\ref{tab:cat} (last column)
while  a cross-correlation with the
2MASS All-Sky Catalog of Point Sources (Cutri et al.~\cite{c:03})
resulted into 68\% X--ray matches with 2MASS counterparts.

Within the 14$'$ radius of the X--ray field of view, the 
2MASS catalog (Cutri et al. 2003) list 26,413 stars, translating into a 
surface density $\mu\sim1.19$ $\times$ 10$^{-2}$~sources~arcsec$^{-2}$.
This corresponds to 0.6 sources within each 
error region of 4$''$ radius. Therefore it is likely that a large 
number of the 2MASS counterparts are just random 
coincidences, as suggested also by the fact that several X--ray sources are
positionally associated with more than one infrared counterpart.
At the spatial resolution of XMM-Newton, stellar confusion in the
Galactic plane prevents from unambiguously associating
infrared sources with the X--ray ones. Thus, we will not discuss further
the possible association with 2MASS sources. 
On the other hand, for the two brightest HD stars, we estimate a probability
of chance coincidence around 0.02\%. This, together with the measured log(f$_{X}$/f$_{V}$),
confirm the real association of the brightest stars with the X--ray sources.

Among the brightest X--ray sources, for which a spectral analysis has
been possible, few can be firmly identified with known objects.
Source~2 in Table~\ref{tab:cat} is a transient  discovered with
EPIC in an XMM-Newton observation performed on 12 March 2003,
pointed on the composite SNR G0.9+0.1 (Sidoli \& Mereghetti 2003;
Sidoli et al. 2004). The observed flux during the discovery
observation was 3.7$\times$10$^{-12}$~erg~cm$^{-2}$~s$^{-1}$
(2--10 keV), almost a factor of 2 higher than that measured in
February 2005, while the spectral parameters remained constant,
within the uncertainties. The detection of the transient in 2005
could indicate that we are observing a second outburst from the
source, or that the source is still in outburst since 2003. 
Source~4 is the pulsar wind nebula in the supernova remnant 
G0.9+0.1 (Mereghetti et al.~\cite{msi:98}). 
The spectral parameters are similar to those
measured during previous observations with \xmm\ (Porquet et
al.~2003; Sidoli et al.~2004). Source~6 is positionally
coincident with source~80 in the ROSAT catalog of the GC 
sources (Sidoli et al.~2001b). Compared with the ROSAT
observation, it displays variability.

Five faint X--ray sources have been detected only above 5~keV. 
While it is somehow expected that the faintest sources be more distant
and absorbed, on the other hand it is remarkable that they
are observed only above 5~keV, meaning that their spectrum is both
truly hard and absorbed. Few of these faint hard sources 
could be background AGNs. From the Log(N)--Log(S) measured in the
energy range 5--10~keV (Baldi et al. 2002) with \xmm, the expected
number of hard sources with flux larger than
5$\times$10$^{-14}$~erg~cm$^{-2}$~s$^{-1}$ is 5--11
sources~deg$^{-2}$, which translates into 1--2 extragalactic hard
X--ray sources in the \xmm\ field of view. The remaining sources
are probably CVs located close to the GC distance
(their fluxes translate into luminosities in the range
$\sim$10$^{32}$--10$^{33}$~erg~s$^{-1}$).


\section{Discussion and Conclusions}
\label{sect:discussion}

In Sidoli et al. (\cite{s:01}) we reported the discovery of a new
X--ray source in the direction of the Sgr~D region, \src.
Our new  \xmm\ observation allows to resolve it into two
sources (sources 3 and 12 in Table~\ref{tab:cat}), 
with a brighter ``main'' source contributing almost 80\% of the
source flux in the 2--10~keV energy range.

The fainter source is harder (detected only above 5~keV) 
than the ``main'' one. A possible optical counterpart is 
the star 0600-28834001 of the USNO-A2.0 catalog (B=18.1, R=13.4),
which is listed as [RHI84]10-672  in the   Raharto et al. (1984) catalog of M-type stars.
The derivation of log(f$_{X}$/f$_{V}$) is highly uncertain, but
assuming, e.g., a blackbody emission at kT$\sim$1~keV,
absorbed with a column density of 10$^{24}$~cm$^{-2}$, the 5--10 keV
flux translates into a 0.3-3.5 keV flux $\sim$4$\times$10$^{-11}$~erg~cm$^{-2}$s$^{-1}$ (corrected for the
absorption),
and to a log(f$_{X}$/f$_{V}$)$\sim$1.8, clearly not stellar.
Thus, the hardness of the X--ray emission excludes a coronal emission for the fainter source.

The refined sky position of the brighter source allows to reject 
all the possible associations
discussed in  Sidoli et al. (\cite{s:01}).
The \sax\ spectrum was affected by a high interstellar absorption,
$N_{\rm H}$$\sim$$10^{23}$~cm$^{-2}$, suggesting that the source
is probably  located at the GC distance (in this case the
luminosity in the 2--10 keV energy band is
$\sim$10$^{34}$~erg~s$^{-1}$). 
A strong Fe~K line was present (with a line centroid of $6.62\pm0.30$~keV),
and a good fit was obtained both with a power-law plus a Gaussian line,
and with a hot thermal plasma model with a temperature, kT, of $6^{+35} _{-4}$~keV. 
Thus, the \sax\ spectrum was consistent with both thermal and
non-thermal models.

The significantly better statistics of the \xmm\ spectrum and the
smaller uncertainties in the spectral slope, favor a non-thermal
nature for the X--ray emission of the ``main'' source. A hard
power-law ($\Gamma \sim$ 1.4) is a good fit to the data, with an
iron line and a high photoelectric absorption (\nh =
10--20~$\times$10$^{22}$~cm$^{-2}$). The absorption is probably
not intrinsic, since the source is located within about 1~degree
from the direction of the Galactic center. Thermal models do not
fit the X--ray spectrum as well, and result in very high
temperatures (for example, a thermal plasma should be hotter than
15~keV). Among the thermal models tried, the blackbody is the best
in fitting the spectrum, but results in a high temperature
($\sim$2~keV) and in an emitting region of less then 0.1~km at the
galactic center distance. Thus, the X--ray spectral shape favors a
non-thermal nature for the X--ray emission.

The X--ray emission appears to be stable; the ``main'' source has
been detected at large off-axis angle during two previous \xmm\
observations performed in September 2000 and March 2003 (both
pointed at the SNR G0.9+0.1). In both occasions, \src\ did not
show evidence for any strong flux variability. Moreover, the total
emission from ``main'' plus ``faint'' source, is compatible with
that observed with  \sax\ in 1997 (see Fig.~\ref{fig:sax}).

These properties are suggestive of three possibilities: a
binary system containing a compact object, a background AGN, or
reflection from a molecular cloud core (e.g., similar to the
X--rays emission and fluorescent iron line produced from the
molecular cloud Sgr~B2; Revnivtsev et al.~\cite{r:04}). 
This third
possibility, already discussed in Sidoli et al.~\cite{s:01}, 
seems now unlikely, based on the high spatial
resolution of the \xmm\ observation. The compact cores contained
in the giant molecular cloud Sgr~B2, for example, are about 1~pc
in size (Lis \& Goldsmith~\cite{lg:91}), while the \xmm\ spatial
resolution (FWHM$\sim$6$''$)
allows us to
exclude a source with a size larger than $\sim$0.25~pc at a
distance of 8.5~kpc.

The shape of the X--ray spectrum and the parameters of the iron
line are consistent with a background AGN. It should be a
nearby object, since the Fe line is not red-shifted. Assuming a
distance of 5~Mpc, the 2--10~keV unabsorbed flux corresponds to a
luminosity of $\sim$3.4$\times$10$^{39}$~erg~s$^{-1}$, which is
quite low, but still compatible with a low-luminosity Seyfert
galaxy (Terashima et al.~\cite{t:02}). Note that no radio
counterpart is present in the NED catalogue within 30$''$ of the
X--ray position, and \src\ does not show evidence for X--ray
variability on timescales of years, while X--ray temporal
variability and presence of radio emission are typical properties
of AGNs.

The X--ray spectral properties of \src\ are reminiscent of the
soft gamma-ray sources discovered with the INTEGRAL
satellite (see e.g., Kuulkers~\cite{k:05} for a review). Several
of their X--ray counterparts display hard and heavily absorbed
spectra, together with intense fluorescent Fe line emission,
indicative of dense gaseous envelopes around the compact object,
illuminated by the central source. In few of them, the association
with OB optical counterparts and the detection of X--ray
pulsations, suggest that they are highly absorbed HMXRBs, not
detected in previous surveys at soft X--rays. The derived
luminosity of these INTEGRAL sources is around
10$^{36}$~erg~s$^{-1}$, although there is a large uncertainty in
the distance estimates, and  the true luminosity could be much
less than this.
\src\ lies in the direction of SgrD molecular cloud, near to SgrB2,
which is
an important site of star formation, so it is not unlikely that
\src\ is indeed a HMXRB.
The low X--ray luminosity suggests that it belongs to a class of
massive X--ray binaries with low persistent emission (in the range
10$^{34}$--10$^{35}$~erg~s$^{-1}$), wind-accreting and with no
outbursts (e.g., 4U~2206+54, Masetti et al.~\cite{m:04}). On the
other hand, these sources typically show temporal variability on
different timescales (sometimes with flares), which has not been
observed in \src\ (perhaps because of the poor coverage). However,
wind-fed HMXRBs are usually quite stable X--ray emitters on long
timescales (months or years). Low luminosity wind-accreting
neutron stars has been predicted by Pfahl et al. (2002), who proposed
that most of the faint sources detected in the $Chandra$ survey 
of the GC (Wang et al. 2002) could
be of this kind. 
A search for hard unidentified sources from ROSAT PSPC observations seems
to confirm that a new class of fainter wind-fed X--ray binaries
exists in our Galaxy (Suchkov \& Hanisch 2004).

Other kinds of galactic X--ray binaries, containing neutron stars
or black-holes, seem to be unlikely; the luminosity
($\sim$10$^{34}$~erg~s$^{-1}$) suggests an object in quiescence:
but low-mass X-ray binaries (LMXRBs) in quiescence (soft X--ray
transients) typically have much softer spectra (blackbody
temperatures $\sim$0.1--0.3~keV; e.g., Verbunt \&
Lewin~\cite{vl:04}), while black-hole X--ray novae in quiescence
have much lower luminosities (Kong et al.~\cite{k:02}).

In conclusion, among the different hypotheses discussed above, the
spectral shape (hard, non--thermal), X--ray luminosity, the
presence of Fe line emission, seem to favor 
a low luminosity HMXRB.


\begin{acknowledgements}
Based on observations obtained with XMM-Newton, an ESA science
mission with instruments and contributions directly funded by ESA
member states and the USA (NASA). We thank Giovanna Giardino,
Nicola La~Palombara and Silvano Molendi for useful discussions.
This research has made use of the SIMBAD database, operated at
CDS, Strasbourg, France. This publication makes use of data
products from the Two Micron All Sky Survey, which is a joint
project of the University of Massachusetts and the Infrared
Processing and Analysis Center/California Institute of Technology,
funded by the National Aeronautics and Space Administration and
the National Science Foundation.
The \xmm\ data analysis is supported by the Italian Space Agency (ASI),
through contract ASI/INAF I/023/05/0.

\end{acknowledgements}

\onecolumn

\clearpage

\begin{landscape}

\begin{table*}[htbp]
\begin{center}
  \caption{The XMM-Newton Catalogue of sources in the region of \src.
The statistical error in the source positions varies 
in the range 0.5--0.8$''$. 
Count rates (in units of ks$^{-1}$) are MOS1+MOS2+pn rates
(see text) observed in the following energy ranges: S =
0.5--2~keV, M = 2--5~keV, H = 5--10~keV. The hardness ratios are
defined as follows: HR1 = (M-S)/(M+S) and HR2 = (H-M)/(H+M). For
the brightest sources (the first seven), the observed fluxes
reported here, in units of 10$^{-13}$~erg~cm$^{-2}$~s$^{-1}$, have
been directly obtained from the spectral fits (see
Table~\ref{tab:other}). For the fainter sources, observed fluxes
(2--10~keV) have been calculated as reported in the text, applying
a conversion factor to the  combined EPIC rate (M+H). For
sources detected only in the soft band (S) or only in the hard
band (H), the fluxes (not corrected for absorption) in the
0.5--2~keV or in the 5--10~keV band have been derived 
and marked with
``S'' or ``H'' respectively. A search within 4$"$ from the source
positions in the SIMBAD data base resulted in the positional
matches listed in the last column (Notes)} \label{tab:cat}
    \begin{tabular}[c]{|l|l|l|c|c|c|l|l|l|l|}
\hline
Source      &    R.A.     & Dec.           &  S                 &     M                     &  H                         &  HR1                               &    HR2    &                  Flux  & Notes$^{a}$ \\

 ID         &    (J2000)  & (J2000)       &                    &                           &                            &                                    &           &                        &       \\
\hline
       1&      267.200047&     -28.211022&         15.14$\pm{     0.67}$&      23.76$\pm{    0.92}$ &       5.99$\pm{     0.47}$ &       0.22$\pm{    0.03}$  &     -0.60$\pm{    0.03} $& 5 & \\
       2&      266.817367&     -28.180065&         2.38$\pm{      0.49}$&      66.02$\pm{    2.27}$ &       29.30$\pm{    1.44}$ &       0.93$\pm{    0.01}$  &     -0.39$\pm{    0.03} $& 20 & Transient (1) \\
       3&      267.070445&     -28.130656&             $-$                &      5.60$\pm{     0.38}$ &       10.87$\pm{    0.46}$ &       1.0                  &      0.32$\pm{    0.04} $& 6.8 & ``main'' source \src   \\
       4&      266.845425&     -28.151610&             $-$                &      20.60$\pm{    1.28}$ &       19.03$\pm{    1.14}$ &       1.0                  &     -0.04$\pm{    0.04}$& 30 & G0.9+0.1  PWN \\
       5&      267.242974&     -28.239226&           7.55$\pm{      0.59}$&      6.82$\pm{     0.67}$ &       2.38$\pm{     0.36}$ &      -0.05$\pm{    0.06}$  &     -0.48$\pm{    0.07}$ & 2 & \\
       6&      266.878546&     -28.229876&           16.66$\pm{     0.99}$&      0.84$\pm{     0.47}$ &       $-$                  &      -0.90$\pm{    0.05}$  &      -1.0                & 0.4 & SBM80 (2); variable \\
       7&      267.149164&     -27.938457&           5.16$\pm{      0.53}$&      5.53$\pm{     0.66}$ &       1.31$\pm{     0.30}$ &       0.03$\pm{    0.08}$  &     -0.62$\pm{     0.08}$& 1.4 & \\
       8&      267.017602&     -28.245968&           0.47$\pm{      0.23}$&      3.11$\pm{     0.45}$ &       5.98$\pm{     0.49}$ &       0.74$\pm{    0.12}$  &      0.32$\pm{     0.07}$& 4 & \\
       9&      267.058541&     -28.272493&           $-$                  &      3.37$\pm{     0.51}$ &       5.39$\pm{     0.52}$ &       1.0                  &      0.23$\pm{     0.08}$& 4 & \\
      10&      267.245308&     -28.188918&           4.45$\pm{      0.42}$&      0.96$\pm{     0.30}$ &        $-$                 &      -0.65$\pm{     0.10}$ &     -0.62$\pm{     0.26}$& 0.5 & \\
      11&      267.202258&     -28.016414&         7.89$\pm{      0.58}$&       $-$                 &        $-$                 &      -1.0                  &       $-$                & 0.2 (S) & HD~316290 \\
      12&      267.081919&     -28.124028&             $-$                &      0.49$\pm{     0.21}$ &       3.15$\pm{     0.31}$ &       1.0                  &      0.73$\pm{     0.10}$& 2 (H) & faint source near \src; \\
       &             &          &                                    &                           &                            &                            &                          & & [RHI84] 10-672;sp.type M6 (3)   \\
      13&      267.071020&     -28.262540&           3.17$\pm{      0.40}$&      1.70$\pm{     0.43}$ &       $-$                  &      -0.30$\pm{     0.13}$ &     -1.0& 0.8 & \\
      14&      267.118308&     -28.158466&           1.45$\pm{      0.22}$&      0.85$\pm{     0.22}$ &       $-$                  &      -0.26$\pm{     0.14}$ &     -1.0 & 0.4 & \\
      15&      267.212103&     -28.245925&           5.32$\pm{      0.49}$&       $-$                 &        $-$                 &      -1.0                  &        $-$               & 0.2 (S) & HD~161824  \\
      16&      267.162129&     -28.176110&           0.31$\pm{      0.15}$&      0.52$\pm{     0.23}$ &       0.78$\pm{     0.18}$ &       0.25$\pm{     0.31}$ &      0.20$\pm{     0.24}$& 0.5 & \\
      17&      266.943950&     -28.150689&             $-$                &       $-$                 &       1.33$\pm{     0.26}$ &       $-$                  &      1.0                 & 0.8 (H) & \\
      18&      266.884630&     -28.148614&             $-$                &       $-$                 &       2.17$\pm{     0.39}$ &       $-$                  &      1.0                 & 1.4 (H) & \\
      19&      267.215469&     -28.180396&           1.27$\pm{      0.23}$&       $-$                 &       $-$ &      -1.0                  &       $-$                & 0.04 (S) & \\
      20&      267.170685&     -28.026434&             $-$                &      0.58$\pm{     0.29}$ &       0.69$\pm{     0.21}$ &       1.0                  &     0.091$\pm{     0.29}$& 0.5 & \\
      21&      267.144386&     -27.926348&           1.64$\pm{      0.35}$&      0.98$\pm{     0.48}$ &       $-$ &      -0.25$\pm{     0.25}$ &     -0.56$\pm{     0.40}$& 0.4 & \\
      22&      267.021045&     -28.117880&             $-$                &         $-$               &       1.69$\pm{     0.22}$ &         $-$                &       1.0                & 1.1 (H) & \\
      23&      266.951207&     -28.065710&             $-$                &         $-$               &       1.83$\pm{     0.27}$ &         $-$                &       1.0                & 1.2 (H) & \\
      24&      267.129677&     -28.197335&             $-$                &         $-$               &       0.87$\pm{     0.19}$ &         $-$                &       1.0                & 0.5 (H) & \\
      25&      266.982671&     -28.052198&           1.85$\pm{      0.27}$&         $-$               &       $-$                  &      -1.0                  &       $-$                & 0.06 (S) & \\
      26&      267.103188&     -28.222073&            $-$                &         $-$               &       0.87$\pm{     0.29}$ &         $-$                &       1.0                & 0.5 (H) & \\
      27&      267.169872&     -28.306960&           1.18$\pm{      0.32}$&         $-$               &        $-$                 &      -1.0                  &       $-$    & 0.04 (S) & \\
      28&      267.044643&     -28.308560&           3.01$\pm{      0.40}$&         $-$               &        $-$                 &      -1.0                  &       $-$    &  0.09 (S)&  TYC2 929  (4) \\
      29&      267.056520&     -28.045786&           $-$                  &         $-$               &       0.86$\pm{     0.17}$ &      $-$                   &       1.0    & 0.5 (H) & \\
      30&      267.057854&     -28.296586&          1.61$\pm{      0.31}$&         $-$               &        $-$                 &      -1.0                  &       $-$    & 0.05 (S) & \\
      31&      267.303578&     -28.077256&          1.42$\pm{      0.26}$&         $-$               &        $-$                 &      -1.0                  &       $-$    & 0.05 (S)& \\
\hline
\end{tabular}
\end{center}
\begin{small}
  $^{a}${Numbers in parentheses are the following references: (1) Sidoli \& Mereghetti \cite{s:03}; (2) Sidoli, Mereghetti \& Belloni \cite{sbm:01}; (3) Raharto et al., \cite{r:84}; (4) Hog et al. \cite{h:00}
} \\
\end{small}
\end{table*}

\end{landscape}

\end{document}